\magnification=1200
\input epsf
\null\bigskip
BIOLOGY AND NONEQUILIBRIUM:

REMARKS ON A PAPER BY J.L. ENGLAND.
\bigskip\bigskip
\centerline{by David Ruelle\footnote{$\dagger$}{Math. Dept., Rutgers University, and 
IHES, 91440 Bures sur Yvette, France. email: ruelle@ihes.fr}.}
\bigskip\bigskip\bigskip\bigskip\noindent
	{\leftskip=2cm\rightskip=2cm{\sl Abstract:}
\medskip\noindent
This note analyzes the physical basis of J.R. England's paper ``Statistical physics of self-replication.''  [J. Chem. Phys. {\bf 139}, 121923(2013)].  We follow England's use of time-reversal symmetry but replace stochastic by deterministic dynamics, and introduce a definition of metastable states based on equilibrium statistical mechanics.  We rederive England's detailed balance relation and obtain another similar relation which appears more natural and remains valid for quantum systems.  The detailed balance relations are based on serious physical ideas, and either of them can be used for England's biological discussion.  This biological discussion does of course deserve further scrutiny.\par}
\vfill\eject
\noindent
{\bf 1. Introduction.}
\medskip
	A recent paper by J.L. England [3] uses non-equilibrium statistical mechanics to predict thermodynamic limitations for biological replication processes.  The conceptual structure of England's paper is such that it permits a serious analysis based on credible physical principles.  In this note we go through a discussion of such physical principles, but unlike [3] we shall not use stochastic dynamics.  We shall aim at being more general (allowing quantum mechanics) and more specific in the conditions of applicability.  \big[Papers related to [3] are [6] and [7], but I prefer the clean conceptual structure of [3]\big].  To state things briefly we shall obtain a variant $(3)$ of England's relation $(1)$ below, equivalent for biological applications but valid also for quantum systems.  The derivation of the relation $(3)$ should make it clear that it is based on general physical principles, and free from ad hoc assumptions.
\medskip
	England's analysis starts with a general detailed balance relation $(1)$ for a system $M$ in contact with a bath at inverse temperature $\beta$.  One considers the conditional probability $\pi(I\to II)$ that the system $M$ is initially in the macrostate $I$ and goes after time $\tau$ into a macrostate $II$.  The reverse probability is $\pi(II\to I)$.  In one intended application, $I$ describes a one-bacterium state which, after cell division, goes over into a state $II$ with two bacteria.  As to the bath it consists of water at inverse temperature $\beta$, with some fixed pressure, containing nutrients, etc.  The detailed balance relation (equation $(8)$ in [3]) is
$$	\beta\langle\Delta Q\rangle_{I\to II}+\ln\Big[{\pi(II\to I)\over\pi(I\to II)}\Big]+\Delta S_{int}\ge0\eqno{(1)}   $$
where $\Delta Q$ is the amount of heat released into the bath and $\Delta S_{int}$ is the increase in internal entropy of $M$.  The meaning of the asymmetric average $\langle\cdots\rangle_{I\to II}$ will be discussed in Sections 4. and 5. below.
\medskip
	The application of $(1)$ to biological replication assumes an exponentially growing number of replicators: $n(t)=n(0)\exp((g-\delta)t)$ where $g$ is a rate of generation and $\delta$ a decay rate.  The formation of a new replicator changes the internal entropy of the system of replicators by $\Delta s_{int}$ while putting an amount $\Delta q$ of heat into the bath.  From $(1)$ one obtains the relation (equation $(10)$ in [3]):
$$	\Delta s_{tot}\equiv\beta\Delta q+\Delta s_{int}\ge\ln[g/\delta]\eqno{(2)}   $$
Note that $g$ and $\delta$ appear here as the ratio $g/\delta$.
\medskip
	From $(2)$ various consequences for the efficiency of replication are derived [the replicator that dissipates more heat has the potential to grow faster, etc.].  For a discussion of these consequences we must refer to [3].
\medskip
	Our study of $(1)$ in the present paper will follow [3], but use deterministic rather than stochastic dynamics.  We shall describe macrostates as unions of metastable states defined according to the principles of equilibrium statistical mechanics.  As in [3] detailed balance will be based on time reversal symmetry of the basic evolution equations.  We shall recover $(1)$ in the deterministic case, but also derive another result:
$$	\beta\langle\Delta Q\rangle+\ln\Big[{\pi^*(II\to I)\over\pi(I\to II)}\Big]+\Delta S_{int}\ge0\eqno{(3)}   $$
where $\langle \cdots\rangle$ is a natural symmetric average obtained from our description of metastable states and $\pi^*(II\to I)$ is an estimate of the probability of $II\to I$.  [Using the probability $\pi(II\to I)$ of the exact time reversed process of $I\to II$ would give equality in $(3)$].  We assume here that $I\to II$ is an observed process, usually irreversible, so that the exact time-reversed process $II\to I$ is physically unobservable.  Very crude estimates $\pi^*(II\to I)$ of $\pi(II\to I)$ are obtained in [3] based on realistic processes.  These very crude estimates from above turn out to be useful because $\pi^*(II\to I)$ occurs in $(3)$ through its logarithm.
\medskip
	The proof of $(3)$ extends naturally to quantum systems (see Section 7) while the proof of $(1)$ in [3] does not, due to the use of stochastic dynamics, and the fact that $\langle\cdots\rangle_{I\to II}$ has no quantum equivalent.  Note that we are still far from a rigorous treatment of $(1)$ or $(3)$ but the use of deterministic dynamics and of a precise definition of metastable states permit a clearer understanding of the limitations of the arguments used.
\medskip
	The relation $(3)$ can be used instead of $(1)$ to derive $(2)$ and justify the biological discussion in [3].  Furthermore, $(3)$ has the advantage of holding also for quantum systems, and permits perhaps a more precise assessment of the applicability of $(2)$.
\bigskip\noindent
{\bf 2. Preliminaries.}
\medskip
	In his discussion of the detailed balance relation $(1)$, England refers to G.E. Crooks [1] who uses classical mechanics to describe the system $M$ of interest and stochastic dynamics for the bath.  The use of stochastic dynamics (Langevin's equation) is traditional and has proved very useful in nonequilibrium statistical mechanics (see for instance [2]), but it limits the understanding of what is physically going on (see [4], [9] for more modern views).  We prefer thus an approach based on deterministic dynamics.  Our discussion will be mainly classical but we shall indicate how to extend it to quantum systems (Section 7).  Indeed, the questions considered here involve chemistry, for which quantum mechanics is important.
\medskip
	The problems discussed in this note pertain to nonequillibrium statistical mechanics.  While there is a serviceable theory of nonequilibrium close to equilibrium (see [2]), it does not apply here because the life processes we are interested in are far from equilibrium.  Note in particular that $(1)$ involves entropy, and there is no general useful definition of entropy far from equilibrium (see for instance [5] and [10] for a discussion of this question).  Here we are saved by two facts:
\medskip
	(F1)  {\bf Description of metastable states.}
\medskip
	Entropy and other thermodynamic variables can be defined for metastable chemical systems.  The definition is imprecise but the imprecision is small for long-lived metastable states (think of diamond or a $O_2+2\,H_2$ mixture at room temperature and pressure).  Metastable states can be described by equilibrium statistical mechanics: take the usual ensembles and restrict them to a dynamically almost isolated region of phase space.
\medskip
	(F2)  {\bf Time reversal and detailed balance.}
\medskip
	The dynamics of the transition $I\to II$ from a macrostate $I$ to a macrostate $II$ may be very difficult to analyze.  But if the dynamics is given by a time-reversal symmetric Hamiltonian, we have control  over the specific quantity $\pi(II\to I)/\pi(I\to II)$ where $\pi(I\to II)$ is the probability of the transition $I\to II$ after time $\tau$, and $\pi(II\to I)$ is the probability of the reverse transition (this is detailed balance).
\medskip
	Our discussion below will be based on (F1), (F2), and we also assume:
\medskip
	(F3)  {\bf Short equilibration times for  the metastable components of $I,II$.}
\medskip
	The equilibration times (thermalization, etc.) for the metastable components of I and II should be short compared with the transition time $\tau$.
\medskip
	We are interested in a system $M$ in a bath.  We shall first think of $M$ as a single molecule which can be in a metastable state $I$ or $II$.  Then we shall consider more general situations where $M$ is for instance a bacterium or two bacteria.  It is physically desirable to surround $M$ by a region $V$ and to let a variable $X$ describe the positions and velocities of $M$ and the bath molecules inside of $V$.
\medskip
	For simplicity we assume short-range forces and choose a region $W\supset V$ such that the inside of $V$ doesn't interact with the outside of $W$.  We let $Y$ describe the position of molecules in $W\backslash V$.\medskip
	Suppose that the bath occupies a large region containing $W$.  Because the bath is large, the effect of changes inside $W$ (such as the release of heat) dissipates rapidly, leaving the region $W\backslash V$ essentially unaffected.
\medskip
	[Unfortunately, the choice of the small regions $V,W$ in a large bath cannot be done neatly in the quantum case due to noncommutativity].
\bigskip
\centerline{\epsfbox{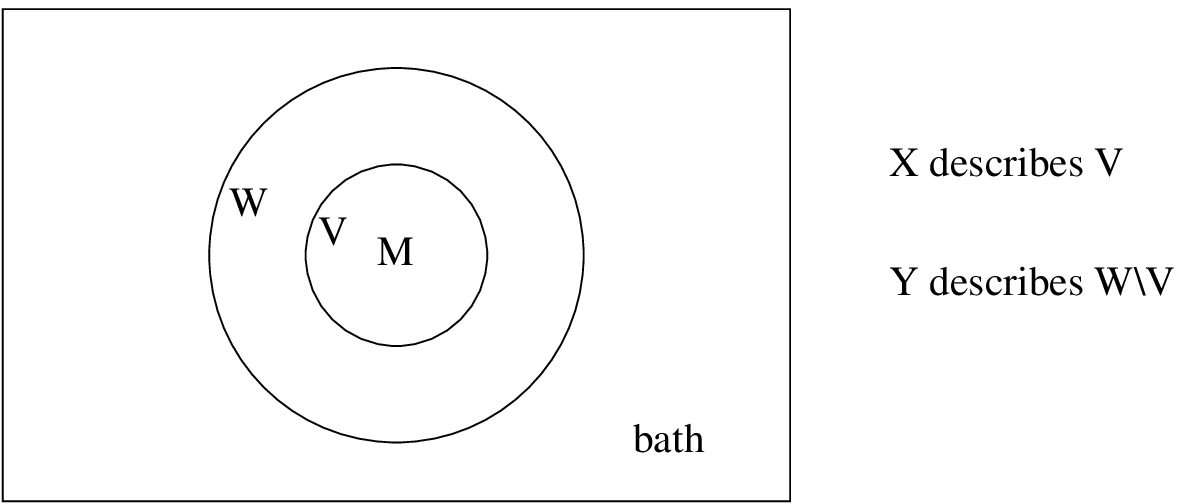}}
\bigskip\noindent
{\bf 3. Estimating the ratio $\pi(II\to I)/\pi(I\to II)$ when $M$ is a single molecule.}
\medskip
	Let $P$ be the phase space describing a body of $H_2O$ (bath) in a large container, plus a molecule $M$ fixed or restrained in the middle of the container.  If we fix the energy $E$ for a suitable Hamiltonian $H(\Omega)$ we see that the volume element $\delta(H(\Omega)-E)\,d\Omega$ in $P$ is preserved by the Hamiltonian time evolution [we may assume that the volume element is ergodic, if not choose a large ergodic component].  We remark that $\delta(H(\Omega)-E)\,d\Omega$ is after normalization the microcanonical ensemble of equilibrium statistical mechanics, for which we shall use the notation ${\bf d}\Omega$.  We suppose here that $H(\Omega)$ has the usual form of kinetic plus potential energy and is thus invariant under time reversal (replacing the momentum ${\bf p}$ by $-{\bf p}$).
\medskip
	As described earlier we surround the molecule $M$  by a set $V$ and a larger set $W$.  We obtain a  statistical description of what is inside $V$ (the molecule $M$ and some water molecules) by integrating over the degrees of freedom outside $V$.  For this it is desirable to replace the above microcanonical ensemble by the canonical ensemble $Z^{-1}\exp(-\beta H(\Omega))\,d\Omega$ for a suitable inverse temperature $\beta$.  The fact that the microcanonical and the canonical ensembles yield the same results inside $V$ in the limit of a large bath is physically reasonable but we shall not try to prove it \big[it would fall in the category of results known as {\it equivalence of ensembles}, see [8]\big].  It is desirable to use the microcanonical ensemble ${\bf d}\Omega$ when discussing time-reversal symmetry, and the canonical ensemble when discussing the system inside $V$.  When we use the classical canonical ensemble it is convenient to ignore momenta (which play a trivial role) and to replace the phase space (positions and momenta) by the configuration space (positions only).  If $X$ describes a point of the configuration space inside $V$ and $Y$ the configuration in $W\backslash V$, the configuration (or potential) energy corresponding to $X$ is of the form $U(X|Y)$ because the molecules outside of $W$ do not interact with those in $V$.  Using the canonical ensemble we find that the probability distribution over the configurations inside $V$ is of the form
$$	[\int dYZ_Y^{-1}\exp(-\beta U(X|Y))]dX\qquad{\rm with}\qquad Z_Y=\int\exp(-\beta U(X|Y))dX\eqno{(4)}   $$
where we have integrated over all positions outside $V$, keeping only the integration over $Y$ explicit.  As indicated above we believe that we would obtain the same distribution $(4)$ in the limit of a large bath if we started with the microcanonical ensemble.  We may also assume that the probability measure for $Y$ denoted here by $dY$ is unique in the limit of a large bath.  [We may include a factor corresponding to an integral over momentum in the definition of $dX$ so that $dX$ is a phase space rather than configuration space volume element: this will be useful later when we discuss entropy].
\medskip
	The macrostates $I,II$ of the molecule $M$ are here single metastable states $I,II$ corresponding to subsets $P^I,P^{II}$ of the phase space $P$ [with total energy fixed at $E$ if we use the microcanonical ensemble].  Metastability means that the sets $P^I,P^{II}$ are dynamically almost isolated from the rest of phase space.  We assume that $P^I$ and $P^{II}$ correspond to sets $R_Y^I$ and $R_Y^{II}$ of configurations $X$ describing the molecule $M$ and the inside of $V$ [the parameter $Y$  corresponds to a configuration in $W\backslash V$].  Note that there is some imprecision in the definition of $R_Y^I$ and $R_Y^{II}$ since the point $\Omega$ describing a metastable state may escape from $P^I$ or $P^{II}$.  The relation assumed between $P^I,P^{II}$ and $R_Y^I,R_Y^{II}$ implies the time-reversal invariance of $P^I,P^{II}$.   If $\theta_Y^I,\theta_Y^{II}$ denote the characteristic functions of $R_Y^I,R_Y^{II}$, we may use $(4)$ to describe the metastable state $I$ as a probability measure (depending on $Y$) with respect to the variable $X$:
$$	(Z_Y^I)^{-1}\theta_Y^I(X)\exp(-\beta U(X|Y))dX\qquad{\rm with}\qquad Z_Y^I
	=\int_{R_Y^I}\exp(-\beta U(X|Y))dX   $$
and similarly for the metastable $II$.
\medskip
	We consider now a possible rearrangement of the molecule $M$ so that it goes from the metastable state $I$ to the metastable state $II$ in time $\tau$.  This means that under the Hamiltonian dynamics $(f^t)$ a fraction of the points in $P^I$ have ended in $P^{II}$ after time $\tau$.  We assume that the dynamics in $P^{II}$ is rapidly mixing so that $f^\tau P^I$ restricted to $P^{II}$ is evenly spread over $P^{II}$ with some coarse-grained density $\lambda$.  [This is the meaning of our assumption (F3) that the equilibration time of $II$ is short with respect to the transition time $\tau$.  We have $\lambda=1$ for a reversible process while $\lambda<1$ for an irreversible process.]  The probabilities of the transitions $I\to II,II\to I$ after time $\tau$ are thus (using the microcanonical volume element ${\bf d}\Omega$)
$$	\pi(I\to II)=\int_{P^I\cap f^{-\tau}P^{II}}{\bf d}\Omega\bigg/\int_{P^I}{\bf d}\Omega   $$
$$	\pi(II\to I)=\int_{P^{II}\cap f^{-\tau}P^I}{\bf d}\Omega\bigg/\int_{P^{II}}{\bf d}\Omega   $$
where
$$	\int_{P^{II}\cap f^{-\tau}P^I}{\bf d}\Omega=\int_{P^{II}\cap f^\tau P^I}{\bf d}\Omega
	=\int_{P^I\cap f^{-\tau}P^{II}}{\bf d}\Omega   $$
[the assumed invariance of ${\bf d}\Omega$ and $P^I,P^{II}$ under time-reversal symmetry give the first equality, the $(f^t)$-invariance of ${\bf d}\Omega$ gives the second].  Therefore
$$	\pi(II\to I)/\pi(I\to II)=\int_{P^I}{\bf d}\Omega\bigg/\int_{P^{II}}{\bf d}\Omega   $$
\par\noindent
{\bf 4. Proof of $(1)$ for general $M$.}
\medskip
	In this section we discuss a situation where $M$ is more general than a single molecule: its macrostates $I$ and $II$ could be a bacterium or two bacteria.  At the same time we replace the bath of $H_2O$ by a bath containing nutrients, etc.  These changes are mostly harmless from our point of view and do not even force us to change notation.  One point however requires discussion: in the case of a single molecule $M$ we assumed that the macrostates $I$ and $II$ are metastable states and that the time $\tau$ associated with the transition between $I$ and $II$ is long with respect to the equilibration times in the states $I$ and $II$.  In more general situations this assumption is not acceptable.  For instance if $I$ is a state of a bacterium with a fixed number of protein molecules, this need no longer be true in state $II$.  We will therefore accept the fact that the macrostates $I$ and $II$ are collections of metastable substates $Ii$ and $IIj$ satisfying (F3).
\medskip
	We may write $P^I=\cup_iP^{Ii}$, $P^{II}=\cup_jP^{IIj}$, but it is readily seen that the formulas written above for $\pi(I\to II),\pi(II\to I)$, and $\pi(II\to I)/\pi(I\to II)$ remain valid (use time-reversal symmetry).  We can now proceed to prove $(1)$.
\medskip
	We have
$$	\int_{P^I}{\bf d}\Omega=\int Z_Y^{-1}dY\int_{R_Y^I}\exp(-\beta U(X^I|Y))dX^I   $$
$$	=\int Z_Y^{-1}dY\,|R_Y^I|^{-1}\int_{R_Y^I}\exp(\ln|R_Y^I|-\beta U(X^I|Y))dX^I   $$
where $|R_Y^I|=\int_{R_Y^I}dX^I$.  Using similar notation for the macrostate $II$, and the identity $|R_Y^{II}|^{-1}\int_{R_Y^{II}}dX^{II}=1$ we find
$$	|R_Y^I|^{-1}\int_{R_Y^I}\exp(\ln|R_Y^I|-\beta U(X^I|Y))dX^I   $$
$$	=|R_Y^{II}|^{-1}\int_{R_Y^{II}}\exp(\ln|R_Y^{II}|-\beta U(X^{II}|Y))dX^{II}   $$
$$	\times|R_Y^I|^{-1}\int_{R_Y^I}\exp(\ln|R_Y^I|-\ln|R_Y^{II}|-\beta(U(X^I|Y)-U(X^{II}|Y))dX^I   $$
Therefore
$$	\int_{P^I}{\bf d}\Omega\bigg/\int_{P^{II}}{\bf d}\Omega=\langle F\rangle_{II}   $$
where
$$ \langle F\rangle_{II}
	=\int Z_Y^{-1}dY|R_Y^{II}|^{-1}\int_{R_Y^{II}}\exp(\ln|R_Y^{II}|-\beta U(X^{II}|Y))dX^{II}\,F(Y,X^{II}) $$
$$	\bigg/\int Z_Y^{-1}dY|R_Y^{II}|^{-1}\int_{R_Y^{II}}\exp(\ln|R_Y^{II}|-\beta U(X^{II}|Y))dX^{II}   $$
$$	F(Y,X^{II})=|R_Y^I|^{-1}\int_{R_Y^I}\exp(\ln|R_Y^I|-\ln|R_Y^{II}|-\beta(H(X^I|Y)-U(X^{II}|Y))dX^I   $$
$$	=\big\langle\exp(\ln|R_Y^I|-\ln|R_Y^{II}|-\beta(H(\cdot|Y)-U(X^{II}|Y))\big\rangle_I   $$
Note that there is no symmetry between $\langle\dots\rangle_I$ and $\langle\dots\rangle_{II}$: see Remark 5 below for a discussion of this point.
\medskip
	We use the notation $\langle\cdots\rangle_{I\to II}=\langle\langle\cdots\rangle_I\rangle_{II}$ and the convexity of exp to obtain
$$	\pi(II\to I)/\pi(I\to II)=\langle F\rangle_{II}
	=\big\langle\exp(\ln|R_Y^I|-\ln|R_Y^{II}|-\beta\Delta Q)\big\rangle_{I\to II}   $$
$$	\ge\exp\langle\ln|R_Y^I|-\ln|R_Y^{II}|-\beta\Delta Q\big\rangle_{I\to II}
	=\exp(-\Delta S_{int}-\beta\langle\Delta Q\rangle_{I\to II})   $$
where the change in internal entropy is
$$	\Delta S_{int}=\langle\ln|R_Y^I|-\ln|R_Y^{II}|\rangle_{II}   $$
[remember that $R_Y$ is a volume in phase space] and the energy given to the bath is
$$	\Delta Q=U(X^I|Y)-U(X^{II}|Y)   $$
We have essentially recovered a proof of $(1)$, apart from an integration over $dY$ not present in [3].  Remember that the condition (F3) is an essential ingredient of our proof.
\bigskip\noindent
{\bf 5. Remark: asymmetry of $(1)$.}
\medskip
	The quantity $\pi(II\to I)/\pi(I\to II)$ is changed to its inverse when $I$ and $II$ are interchanged.  In the expression
$$	\big\langle\exp(\ln|R_Y^I|-\ln|R_Y^{II}|-\beta\Delta Q)\big\rangle_{I\to II}   $$
this symmetry is preserved but hidden because $\langle\cdots\rangle_{I\to II}$ is not symmetric.  After using the convexity of exp we obtain
$$	\ge\exp\langle\ln|R_Y^I|-\ln|R_Y^{II}|-\beta\Delta Q\big\rangle_{I\to II}   $$
where the symmetry is broken.  The symmetry is also broken in $(1)$: this is why we have $\ge0$ instead of $=0$.  The physical meaning of introducing an asymmetry between $I$ an $II$ by using $\langle\cdots\rangle_{I\to II}$ is not clear to the present author [this applies to the derivation in [3] as well as the one given here].
\bigskip\noindent
{\bf 6. Proof of $(3)$.}
\medskip
	Omitting a superscript $I$ or $II$, let again $\theta_Y$ be the characteristic function of $R_Y$ and write
$$	\rho(X,Y)=\theta_Y(X)e^{-\beta U(X|Y)}\bigg/\int\int_{R_Y}e^{-\beta U(X|Y)}dX\,dY   $$
$$	S=\int\int\big[-\rho_Y(X)\ln\rho_Y(X)\big]dX\,dY   $$
$$	=\ln\int\int_{R_Y}e^{-\beta U(X|Y)}dX\, dY+\int\int\beta U(X|Y)\rho_Y(X)dX\,dY   $$
We have thus
$$	\ln\Big[{\pi(II\to I)\over\pi(I\to II)}\Big]
	=\ln\int\int_{R_Y^I}e^{-\beta U(X|Y)}dX\, dY-\ln\int\int_{R_Y^{II}}e^{-\beta U(X|Y)}dX\, dY   $$
$$	=S^I-S^{II}-\beta\int\int U(X|Y)[\rho_Y^I(X)-\rho_Y^{II}]dX\,dY=-\Delta S_{int}-\beta\langle\Delta Q\rangle   $$
with natural symmetric definitions.
\medskip
	Instead of an inequality obtained via the asymmetric $\langle\cdots\rangle_{I\to II}$ and (the somewhat artificial) use of the convexity of the exponential, we have here an equality.  Note however that $\pi(II\to I)$ is the probability of the inverse of a usually ``irreversible'' process (this inverse process is therefore somewhat unrealistic).  What is done in [3] is to obtain estimates $\pi^*(II\to I)>\pi(II\to I)$ in terms of realistic processes (like rate of degradation of nucleic acids).  We obtain thus $(3)$ finally in the form
$$	\ln\Big[{\pi^*(II\to I)\over\pi(I\to II)}\Big]\ge-\Delta S_{int}-\beta\langle\Delta Q\rangle   $$
As mentioned earlier even a very crude estimate $\pi^*(II\to I)$ yields interesting results because $\pi^*(II\to I)$ occurs as its logarithm in $(3)$.
\bigskip\noindent
{\bf 7. Quantum systems.}
\medskip
	Consider a quantum system with Hamiltonian $H$ acting on the Hilbert space ${\cal H}$, and let $P^I,P^{II}$ be projections in ${\cal H}$.  We assume that $H$ and $P^I,P^{II}$ are invariant under time reversal and that  $P^I,P^{II}$ are almost invariant under time evolution.
\medskip
	The microcanonical ensemble for the energy $E$ corresponds here to the eigenfunctions $\psi$ of $H$ such that $H\psi\approx E\psi$.  We let $P_E$ be the projection on the space spanned by these eigenfunctions.  We have now
$$	\pi(I\to II)={\rm Tr}(P_E(P^Ie^{i\tau H}P^{II})(P^{II}e^{-i\tau H}P^I))/{\rm Tr}(P_EP^{II})   $$
$$	\pi(II\to I)={\rm Tr}(P_E(P^{II}e^{i\tau H}P^I)(P^Ie^{-i\tau H}P^{II}))/{\rm Tr}(P_EP^I)   $$
Writing $P_E^I=P_EP^I$, $P_E^{II}=P_EP^{II}$ we find
$$	{\rm Tr}((P_E^Ie^{i\tau H}P_E^{II})(P_E^{II}e^{-i\tau H}P_E^I))
	={\rm Tr}((P_E^{II}e^{-i\tau H}P_E^I)(P_E^Ie^{i\tau H}P_E^{II}))   $$
$$	={\rm Tr}((P_E^{II}e^{i\tau H}P_E^I)(P_E^Ie^{-i\tau H}P_E^{II}))   $$
where we have used time-reversal symmetry.  Therefore
$$	\pi(II\to I)/\pi(I\to II)={\rm Tr}(P_EP^I)/{\rm Tr}(P_EP^{II})   $$
\par
	We have metastable states localized in some sense in a small region $V$ of the bath and we may replace the microcanonical ensemble by a (grand) canonical ensemble as explained in the classical case.  We obtain
$$	\pi(II\to I)/\pi(I\to II)={\rm Tr}(e^{-\beta H}P^I)/{\rm Tr}(e^{-\beta H}P^{II})   $$
[Due to noncommutativity we cannot here strictly restrict our attention to small regions $V,W$ as in the classical case].  Omitting the superscript $I$ or $II$ we write
$$	\rho=e^{-\beta H}P/{\rm Tr}(e^{-\beta H}P)\qquad,\qquad S=-{\rm Tr}\rho\ln\rho   $$
then
$$	S=\ln{\rm Tr}(e^{-\beta H}P)+\beta\,{\rm Tr}\rho H   $$
Therefore
$$	\ln\Big[{\pi(II\to I)\over\pi(I\to II)}\Big]
	=S^I-S^{II}+\beta\,{\rm Tr}(\rho^I H-\rho^{II}H)=-\Delta S-\beta\langle\Delta Q\rangle   $$
from which the quantum version of $(3)$ follows.  We leave the details to the reader.
\bigskip\noindent
{\bf 8. Loose ends and final remarks.}
\medskip
	Life is fueled by chemical energy but also by photosynthesis.  We can deal with the latter by saying that it creates high energy metastable states and that the rest is chemistry.  At the level of generality of the present discussion it seems that there is not much more that can be said.
\medskip
	We have presented life processes in our discussion as a succession of jumps between quasi-equilibrium metastable states with relaxation to metastability between the jumps.  But note that in our derivation of detailed balance relations no precise description of what happens between states $I$ and $II$ is used: we only need a rough description of $I$ and $II$ and a rough estimate $\ln\pi^*(II\to I)>\ln\pi(II\to I)$.  For this rough estimate the states $I$ and $II$ may indifferently be dead or living bacteria: no subtle difference between the entropy of a living and a dead bacterium plays any role here.
\medskip
	Let us discuss once more the relation between time-reversal symmetry and irreversibility.  The time-reversal symmetry used in proving $(1)$ or $(3)$ holds for the Hamiltonian time evolution on the energy shell (i.e., the microcanonical ensemble, which we take to be ergodic).  Introducing stochastic forces as in [3] or replacing the microcanonical ensemble by another ensemble complicates the time reversal symmetry and the estimation of the very small probabilities $IIj\to Ii$.  If we assume that the process $I\to II$ is ``irreversible'' this implies that the time-reversed dynamics is very unstable so that $\pi(II\to I)$ is hard to estimate precisely.  One uses instead a bigger probability $\pi^*(II\to I)$ based on observable processes.
\medskip
	From a conceptual viewpoint the paper of England, as rediscussed in the present note, is based on general thermodynamics (including distinction between reversible and irreversible processes, Carnot cycles, etc.).  These thermodynamic ideas are complemented by ideas of Hamiltonian dynamics (time-reversal symmetry) and of equilibrium statistical mechanics (metastable states and relation between entropy and probabilities).  The well-known relation between entropy and probabilities appears difficult to exploit for probability statements about biological replication.  The beauty of England's paper [3] is that it succeeds in making such statements, based on time-reversal symmetry and some accessible estimates of reverse processes.  The purpose of the present note has been to clean the theoretical basis of England's arguments, using deterministic dynamics and a definition of metastable states based on equilibrium statistical mechanics.  We have proposed to replace $(1)$ by $(3)$ which holds also for quantum systems, but the biological applications and conclusions remain those presented by J. England.
\vfill\eject\noindent
{\bf References.}
\medskip\noindent
[1] G.E. Crooks.  ``Entropy fluctuation theorem and the non equilibrium work relation for free energy differences.''  Phys. Rev. E {\bf 60},2721(1999).
\medskip\noindent
[2] S.R. de Groot and P. Mazur.  {\it Nonequilibrium thermodynamics.}  Dover, New York, 1984.
\medskip\noindent
[3] J.L. England.  ``Statistical physics of self-replication.''  J. Chem. Phys. {\bf 139},121923 (2013).
\medskip\noindent
[4] G. Gallavotti.  {\it Nonequilibrium and Irreversibility.}  Springer, Cham, 2014.
\medskip\noindent
[5] J.L. Lebowitz.  ``Boltzmann's entropy and time's arrow.''  Physics Today {\bf 46}, No 9,32-38(1993).
\medskip\noindent
[6] R. Marsland and J. England.  ``Thermodynamic expression for non equilibrium steady-state distribution of macroscopic observables.'' arXiv:1501.00238 (2015).
\medskip\noindent
[7] N. Perunov, R. Marsland and J. England.  ``Statistical physics of adaptation.''  arXiv: 1412.1875 (2014).
\medskip\noindent
[8] D. Ruelle.  {\it Statistical Mechanics, Rigorous Results.}  Benjamin, New York, 1969.
\medskip\noindent
[9] D. Ruelle.  ``Smooth dynamics and new theoretical ideas in nonequilibrium statistical mechanics.''  J. Statist. Phys. {\bf 95},393-468(1999).
\medskip\noindent
[10] D. Ruelle.  ``Extending the definition of entropy to nonequilibrium steady states.''  Proc. Nat. Acad. Sci. {\bf 100},3054-3058(2003).
\end